\documentclass[9pt,twocolumn,twoside]{opticajnl}
\journal{opticajournal} 

\setboolean{shortarticle}{true}

\usepackage{lineno}

\usepackage{comment}

\title{Regenerative vectorial breathers in a delay-coupled neuromorphic microlaser with integrated saturable absorber}

\author[1,*]{Stefan Ruschel}
\author[2]{Venkata A. Pammi}
\author[2,3,4]{Rémy Braive}
\author[2]{Isabelle Sagnes}
\author[2]{Grégoire Beaudoin}
\author[5,6]{Neil G. R. Broderick}
\author[5,6]{Bernd Krauskopf}
\author[2]{Sylvain Barbay}

\affil[1]{University of Leeds, Woodhouse, Leeds LS2 9JT, United Kingdom}
\affil[2]{Université Paris-Saclay, CNRS, Centre de Nanosciences et de Nanotechnologies, 10 Bd Thomas Gobert, 91120 Palaiseau, France}
\affil[3]{Université Paris-Cité, 75207 Paris Cedex 13, France}
\affil[4]{Institut Universitaire de France, Paris, France}
\affil[5]{Dodd-Walls Centre for Photonic and Quantum Technologies, New Zealand}
\affil[6]{The University of Auckland, Private Bag 92019, Auckland 1142, New Zealand}

\affil[*]{s.ruschel@leeds.ac.uk}

\begin{abstract}
We report on the polarization dynamics of regenerative light pulses in a micropillar laser with integrated saturable absorber coupled to an external feedback mirror. The delayed self-coupled microlaser is operated in the excitable regime, where it regenerates incident pulses with a supra-threshold intensity --- resulting in a pulse train with inter-pulse period approximately given by the feedback delay time, in analogy with a self-coupled biological neuron. 
We report the experimental observation of vectorial breathers in polarization angle, manifesting themselves as a modulation of the linear polarized intensity components without significant modulation of the total intensity.
Numerical analysis of a suitable model reveals that the observed polarization mode competition is a consequence of symmetry-breaking bifurcations induced by polarization anisotropy. Our model reproduces well the observed experimental results and predicts different regimes as a function of the polarization anisotropy parameters and the pump parameter.
We believe that these findings are relevant for the fabrication of flexible sources of polarized pulses with controlled properties, as well as for neuroinspired on-chip computing applications, where the polarization may be used to encode or process information in novel ways.
\end{abstract}

\setboolean{displaycopyright}{false} 

\begin{document}

\maketitle

The capacity of optical devices to generate pulses of light can be exploited in various applications, ranging from optical communication to optical information processing \cite{McMahonNRP23}. The polarization of light pulses constitutes an additional degree of freedom that can be leveraged for specific applications \cite{Zhao22}. Polarization also introduces new dynamical phenomena, such as vectorial chaos in semiconductor vertical cavity surface emitting lasers (VCSELs) \cite{ScirePRL03}, vectorial self-organization in fiber mode-locked lasers \cite{KrupaO17}, and spontaneous symmetry-breaking or breathing dissipative vectorial solitons in Kerr resonators \cite{XuNC21,XuOL22,HillCP24} or mode-locked fiber lasers \cite{LuoOE20,HuangAPN23}. Orthogonally polarized frequency combs have been predicted numerically in a mode-locked vertical external cavity surface emitting laser (VECSEL) with saturable absorber (SA)\cite{VladimirovOL19}. Moreover, vectorial dissipative solitons have been reported experimentally in VCSELs subjected to polarization selective feedback plus cross-polarized reinjection \cite{MarconiNP15}; coexistence of antiphase bright and dark temporal dissipative solitons was observed in both polarization components.

Here, we focus on regenerative self-pulsation \cite{GarbinNC15,Yanchuk2019} in semiconductor microlasers with integrated SA and delayed optical feedback. These micropillar lasers have been demonstrated experimentally \cite{Pammi2019} to serve as building blocks for spike processing in brain inspired photonic applications\cite{prucnal2017neuromorphic,Shastri2021}. They also have been shown theoretically, by using the polarization degree of freedom, to allow for inhibitory neural dynamics \cite{ZhangOL19} and XOR logical operation \cite{XiangOL20}. In micropillar lasers, the laser emission takes place perpendicular to the device surface as in standard VCSELs and the lateral cavity geometry has circular symmetry. Therefore, the emission polarization is not fixed by the geometry but is, nevertheless, subject to anisotropies due to the crystallographic axes leading to dichroism and birefringence. These effects couple the two orthogonal polarization components observed (usually linear polarizations), and this can lead to instabilities. In the case of solitary microcavity lasers with SA, it was shown theoretically, while neglecting dichroism, that the emission can display intensity and polarization pulsations \cite{ScireOL02}.

Regenerative self-pulsation considered in this Letter occurs below the solitary laser self-pulsing threshold. Physically, when the device is in the excitability regime \cite{Barbay2011,Selmi2014}, its steady state is the laser-off state; however, small perturbations can trigger characteristic excitable pulses, which are calibrated excursions in phase space. If the delayed optical feedback is sufficiently strong, the excitable pulse can regenerate itself after one round trip in the external feedback loop and, hence, give rise to a train of pulses, sometimes referred to as a temporal dissipative soliton \cite{Yanchuk2019}. The interest in this regime for spike processing lies in the manipulation possibilities of the optical pulses offered: they can be written, erased and tweezed in the short term --- giving rise to an optical buffer memory, or to an almost arbitrary pulse sequence generator \cite{Terrien2018a,Ruschel2020limits,PammiThese21}. In the long term, however, any generated pulse pattern must converge to one of the simultaneously stable periodic pulsing regime \cite{Terrien2019}, which is interesting for building smart memories with self-healing properties. Mathematically, the introduction of delayed optical feedback makes the system infinite dimensional and generates a wealth of complex dynamical phenomena \cite{TerrienChaos23}, including locking dynamics on tori and spontaneous symmetry-breaking of pulse timings \cite{TerrienPRE21}. 

We investigate experimentally and numerically the polarization dynamics of a microlaser with SA with delayed optical feedback in its regenerative pulsing regime. Specifically, we report experimental vectorial breathing (polarization) dynamics of regenerative pulse trains. By introducing a suitable model, we recover the experimental result and are able to analyze theoretically the possible dynamical regimes in a realistic range of the polarization anisotropy parameters. We identify and highlight regions of pure polarization breathing with constant total intensity, as well as those with mixed intensity-polarization modulations.

\begin{figure}[!]
	\includegraphics[width=\linewidth]{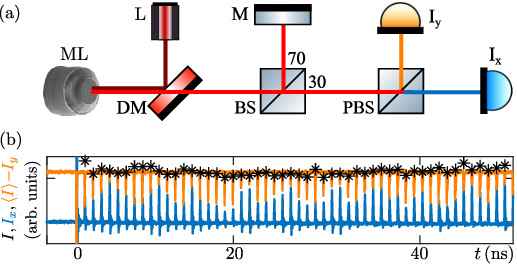}
	\caption{(Preliminary figure) The experimental set-up and observed time-series. (a) Schematic of experimental setup. (b) Experimental vectorial pulse train with modulated linear polarization intensities $I_x$ (blue) and $I_y$ (orange, illustrated by its complement w.r.t. the total intensity $I$), while the peaks of $I$ (black asterisks) remain almost constant. \label{fig:experiment}}
\end{figure}

\emph{Experimental results.}
The experimental setup is depicted in Fig.\ref{fig:experiment}a. A micropillar laser (ML) of 5 micron diameter with an integrated SA is thermally controlled by a Peltier PID loop and optically pumped by a cw laser diode at ~800nm through a microscope objective. The 980nm emission of the ML passes through a dichroic mirror (DM) and is directed via a 70/30 (R/T) beamsplitter (BS) to a high reflectivity mirror (M) that reflects part of the incoming light back into the ML; the roundtrip time is about 8.8ns. In the steady-state regime and in absence of any external input, the system is in the laser-off state. The transmitted part of the light passes through a polarizing beamsplitter (PBS) that separates the $x$ and $y$ polarization components, which are detected by high-bandwitdh photodiodes. To start a pulse train, a single 80ps optical pulse at about the pump wavelength is send to the micropillar laser. This input pulse is produced by a mode-locked Ti:Sa laser (L) with a pulse-picker that allows to select single pulses and to modulate the output repetition period. If the input pulse has sufficient amplitude, it triggers a response pulse in the ML which, for sufficient external cavity feedback strength, produces a vectorial pulse train. As observed in Fig.\ref{fig:experiment}b, such a pulse train repeats in each analyzed polarization component with a repetition period approximately given by the feedback delay duration, as expected. More importantly, a pronounced amplitude modulation appears in each of the polarization components $I_x$ and $I_y$ (illustrated by its complement), while the modulation of the peaks (black asterisks) of the total intensity $I$ remains small. The antiphase dynamics of $I_x$ and $I_y$ reveals the characterizing feature of a regenerative vectorial breather: only the polarization angle, not the intensity of the single component, is modulated after each round trip. This polarization modulation has a period of about 5-6 delay times. Because of pump and detection noise it is difficult to say whether the shown output corresponds to locked dynamics of the quasi-periodic regime. In any case, it is well known that 
locked dynamics in this system exists over a large parameter range \cite{TerrienChaos23}. 

\emph{Mathematical model.}
Our starting point for a polarization-resolved mathematical model is the Spin Flip Model (SFM) for a VCSEL with SA \cite{ScireOL02}. It describes the evolution and interaction of two circular polarization components $F_{\pm}$ of the complex electric field and the corresponding real gain and saturable absorber variables. We consider here a generalization with a delayed optical feedback term that accounts for the effect of the external feedback mirror; this extended spin-flip model is given by
\begin{align}
2\dot F_{\pm}(t) =& \left[(1+i\alpha)G_{\pm}(t)
-(1+i\beta)Q_{\pm}(t)-1\right]F_{\pm}(t) \nonumber\\
&- (\varepsilon_a + i\varepsilon_p) F_{\mp}(t)
+\kappa e^{i\psi}F_{\pm}(t-\tau),\label{eq:F-def}\\
\dot G_{\pm}(t) =& \gamma_{G}(A-G_{\pm}(t)(1+|F_{\pm}(t)|^{2})) \nonumber\\
&-\delta_G(G_{\pm}(t)-G_{\mp}(t)),\label{eq:G-def}\\
\dot Q_{\pm}(t) =& \gamma_{Q}(B-Q_{\pm}(t)(1+a|F_{\pm}(t)|^{2}))\nonumber\\
&-\delta_Q(Q_{\pm}(t)-Q_{\mp}(t)).
\label{eq:Q-def}
\end{align}
Here, time is rescaled to the photon lifetime in the cavity, and the rescaled variables $G_\pm$ and $Q_\pm$ for gain and SA, respectively, represent the population inversions of the carrier reservoirs coupling to the left and right circularly polarized components of the electric field. Polarization anisotropy is represented by the parameters $\varepsilon_a$ and $\varepsilon_p$, which depend not only on the materials used to fabricate the lasers but may also vary as a result of crystal mechanical stress; therefore, their values may vary even for nearby devices on a wafer, due to fabrication induced anisotropies. It is known, however, that in VCSELs usually the linear dichroism is small whereas birefringence is substantially larger, and this explains the mainly linear polarization emission of these devices \cite{MartinRegaladoIEEEJQE97,ExterPRL98,VanDerSande2006,Panayotov2013}. Note that system (\ref{eq:F-def})--(\ref{eq:Q-def}) is invariant under a phase shift over $\pi$ (multiplication by $-1$) applied to only one of the two fields $F_{\pm}$ and changing simulaneously the signs of both $\varepsilon_a$ and $\varepsilon_p$; hence, we may restrict our exploration and model analysis to positive values of $\varepsilon_p$. The parameters $\alpha$ and $\beta$ are the linewidth enhancement factors in the gain and SA regions, and $\gamma_{G,Q}, A, B, a, \delta_{G,Q}$ describe respectively: the relaxation rates of the SA and gain sections; the pump level of the gain; the linear losses of the SA; the saturation parameter; and the reduced spin-flip rates in gain and SA, such that $\delta_{G,Q}=1/2(\gamma_{s_{G,Q}}-\gamma_{G,Q}$) with the spin flip rates $\gamma_{s_{G,Q}}$. Finally, the feedback with roundtrip time delay $\tau$ has strength $\kappa$ and feedback phase $\psi$ between the delayed and instantaneous fields.

Without polarization anisotropies (when $\varepsilon_a=\varepsilon_p=0$), system (\ref{eq:F-def})--(\ref{eq:Q-def}) reduces to the well-known Yamada model of a laser with SA \cite{Yamada1993} in the absence of delayed optical feedback (when $\kappa =0$). Its bifurcation structure has been studied in \cite{Dubbeldam1999} and, when delayed optical feedback is included, the Yamada model shows regenerative self-pulsation that were studied theoretically  \cite{Krauskopf2011,Terrien2018,Ruschel2020limits} and observed experimentally  \cite{Terrien2017a,Terrien2018a}.

\emph{Results of numerical analysis.}
To study the polarization dynamics of such sustained pulse trains we perform numerical simulations of model (\ref{eq:F-def})--(\ref{eq:Q-def}) for the parameter values shown in Tab \ref{tab:1}, if not specified otherwise. They ensure that the system is in the excitable regime \cite{Barbay2011}, so that the off solution of the ML ($F_\pm=0$) is stable and coexists with regenerative self-pulsation. In our simulations, we choose pulse-like initial conditions for the two fields with constant gain and SA variables, which rapidly converge to one-pulse-per-roundtrip regenerative pulses; specifically, this yields either $F_+=F_-$ or $F_+=-F_-$, corresponding to the pulses being exclusively in one linear polarization component $F_x=(F_+ +F_-)/\sqrt{2}$ or $F_y=(F_+ -F_-)/\sqrt{2}$, respectively (with the other being zero). To study the influence of the anisotropy parameters $\varepsilon_a$ and $\varepsilon_p$, we perform systematic numerical simulations in the physically plausible parameter range $\varepsilon_a\ll 1$ and $0<\varepsilon_p\leq 0.5$. Note that, again by a symmetry argument, we can restrict attention to $\varepsilon_a>0$ as well.

\begin{figure}[!]
\includegraphics[width=\linewidth]{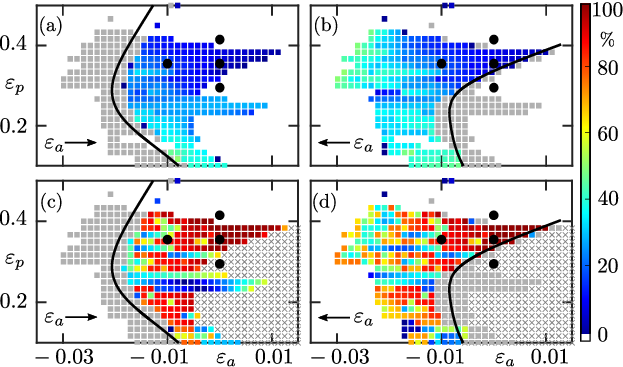}
\caption{Relative modulation of total pulse intensity $I$ (a)--(b) and  pulse intensity $I_x$ (c)--(d) in the $x$-polarization of system (\ref{eq:F-def})--(\ref{eq:Q-def}) as a function of the amplitude and phase anisotropies $\varepsilon_a$ and $\varepsilon_p$ (other parameters as in table \ref{tab:1}). Shown is the range of pulse intensities (difference between maximum and minimum heights of pulses) relative to the observed maximum intensity pulse (in $\%$ according to the color bar). Panels (a) and (c) show the $\varepsilon_a$-up-sweep, and panels (b) and (d) the $\varepsilon_a$-down-sweep, where nonzero results of the respective other sweep are shown as gray squares. Also shown are two torus bifurcation curves (black) bounding the region of nonzero modulation; gray crosses indicate the region where $I_x = 0$.
\label{fig:2Dbif}}
\end{figure}

Figure \ref{fig:2Dbif} shows the results of parameter up- and down-sweeps 
in the amplitude anisotropy $\varepsilon_a$ (horizontal axis) for different values of the phase anisoptry $\varepsilon_p$ (vertical axis). Shown in a color code is the relative variation $(I_\mathrm{max}-I_\mathrm{min})/I_\mathrm{max}$ of the maxima and minima of the pulse heights, in terms of the total intensity $I=|F_+|^2+|F_-|^2$  in panels (a)--(b), and the $x$-polarization component intensity $I_x=|F_x|^2$ in panels (c)--(d). These quantities were determined from a computed time series of model (\ref{eq:F-def})--(\ref{eq:Q-def}) over $100$ roundtrips in the feedback loop, after discarding a long transient of 2000 roundtrips. Panels (a)--(b) show hysteresis between up- and down-sweeps of the total peak intensity modulations observed in the simulations; the region of bistability is indicated by gray squares. We find that the onset of modulation of the total peak intensity is mediated by two torus bifurcation curves (black): one of the $x$-polarized pulse ($F_+=F_-$, $F_y=0$) in panel (a), and a second of the $y$-polarized pulse ($F_+=-F_-$, $F_x=0$) in panel (b). A torus bifurcation curve can be computed with the numerical continuation software package DDE-Biftool \cite{Sieber2014}, and it corresponds to the emergence of a second frequency, which is the modulation frequency of the polarization modes in our case. This is confirmed by the corresponding up- and down-sweeps of the $x$-polarization intensity $I_x$ in panels (c)--(d). 

We observe that the modulation of the total intensity is initially small (as expected), but grows quickly as we move away from the respective torus curve in parameter space; see Fig. \ref{fig:2Dbif}(a)--(b). In the linear polarization basis, this has a more dramatic effect, as is illustrated in Fig. \ref{fig:2Dbif}(c)--(d). Since one of the two polarization fields is zero before the respective torus bifurcation, all modulation of the peak intensity will contribute to that of the pulse height in the respective linear component. Therefore, the maximum variability in the intensity is much higher for the linear component, $F_x$ or $F_y$, compared to that for the circular basis, $F_+$ and $F_-$. On the other hand, in the parameter region $-0.01\leq\varepsilon_a\leq0.05$ and $0.3\leq\varepsilon_p\leq0.4$ the variation of the total peak intensity $I$ remains quite small (less than 10\%), while that in $I_x$ (and, hence, also $I_y = I = I_x$) is substantial. This type of pulse dynamics matches that of our experimental findings. Note further that there is little to no multistability in this parameter region, making it more likely to observe this pulsing regime consistently in the experiment. 

\begin{figure}[!]
\includegraphics[width=0.95\columnwidth]{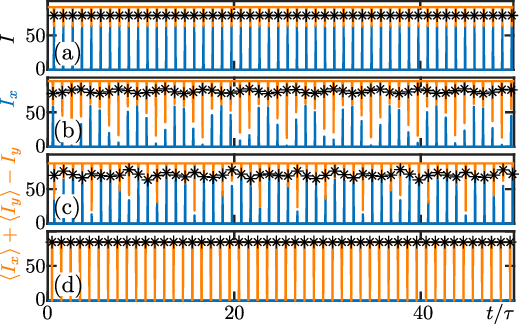}
 \caption{Representative time series of model (\ref{eq:F-def})--(\ref{eq:Q-def}) for anisotropy parameters at the black dots indicated in Fig.\ref{fig:2Dbif}; namely for (a) $(\varepsilon_a,\varepsilon_p)=(0,0.415)$, (b) $(\varepsilon_a,\varepsilon_p)=(0,0.355)$, (c) $(\varepsilon_a,\varepsilon_p)=(-0.01,0.3555)$, and (d) $(\varepsilon_a,\varepsilon_p)=(0,0.295)$ (other parameters as in table \ref{tab:1}). The y-component $I_y$ (orange) is shown as the complement of $I_x$ (blue); compare with Fig.~\ref{fig:experiment}(b). \label{fig:timeseries}}
\end{figure}

Figure \ref{fig:timeseries} shows example time series at the parameter value indicated by black dots in Fig. \ref{fig:2Dbif}. Shown throughout is the $x$-polarization component $I_x$ with the complement of the $y$-polarization component $I_y$, to show how they contribute to the total intensity $I$. Figure \ref{fig:timeseries}(a) shows a solution with fixed ratio of contributions of the $x$- and $y$-components to the total intensity. This corresponds to exact antiphase peak dynamics of the circular polarizations, i.e., one large and one smaller peak that alternate between the circular polarizations from one period to the next. For slightly lower $\varepsilon_p$ as in panel (b), we observe a kind of unlocking, which gives rise to a quite regular modulation of the pulses in the linear polarization components. This time series matches well the experimental observations in Figure \ref{fig:experiment}(b) and, hence, provides an indication of the actual experimental polarization anisotropy. When decreasing $\varepsilon_a$ to move away from the torus bifurcation curve in Fig. \ref{fig:2Dbif}, the modulation of the pulses of $I_x$ and $I_y$ as is illustrated in Fig. \ref{fig:timeseries}(c). When further decreasing the value of $\varepsilon_p$ from that in panel (b), on the orther hand, the torus bifurcation curve is crossed and a stable pulsation regime is entered where the circular component have equal intensity, but a phase shift of $\pi$ between them. Hence, $F_+ = F_-$, and the regenerative pulses have only the $y$-component, with $I_x = 0$, as is the case for the time series in Fig. \ref{fig:timeseries}(d).

\begin{figure}[!]
\includegraphics[width=\columnwidth]{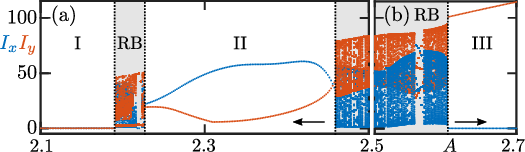}
\caption{
Down-sweep (a) and up-sweep (b) in $A$, both starting at $A = 2.5$ (time series shown in Fig. 3(b), other parameters as in table \ref{tab:1}), showing the pulse trains peak intensity in $I_x$ (blue) and $I_y$ (orange), identifying the range of regenerative vectorial breathers (RB) and types I--III of pulse dynamics (see text).
\label{fig:Asweeps}}
\end{figure}

We close by considering the effect of changes in the pump parameter $A$ (still all below the lasing threshold). Figure~\ref{fig:Asweeps} shows the maximum peak intensities in the linear polarization basis during a down-sweep (a) and an up-sweep (b) in $A$, both starting from the regenerative vectorial breather at $A=2.5$ shown in Fig.~\ref{fig:Asweeps}(b). In either case, it can be tracked over a considerable range, showing that the regenerative vectorial breather regime (RB, shaded) is quite robust with respect to $A$. Moreover, we find three additional regimes: I. laser off (no regenerative pulses); II. regenerative pulses with fixed polarization angle; and III. purely $y$-polarized regenerative pulses; note also the second RB regime near $A = 2.2$. This analysis shows that the pump parameter may serve as an efficient additional knob to control the observed pulsation dynamics, in light of the large, and difficult to master, variations of the anisotropy that can be observed in practice.

\emph{Conclusion.}
We have experimentally observed and modeled vectorial regenerative breathers in the linear polarization components of a delay-coupled neuromorphic micropillar laser with integrated saturable absorber. We focused on polarization mode competition and mapped out its dependence on the parameters characterizing the polarization anisotropy. In this way, we revealed the region of existence of linearized polarization dynamics in the presence of an effectively constant overall intensity. Our results may help enable the fabrication of flexible sources of polarized light pulses with controlled properties, as well as the designing of neuromorphic optical circuits that harnass polarization degrees of freedom to implement artificial neural networks.

\begin{table}[t]
\centering
\caption{\bf Model parameters}
\scalebox{0.75}{
\begin{tabular}{ccccccccccccc}
\hline
$\alpha$ & $\beta $ & $\gamma_G$ & $\gamma_Q$ & $\delta_G$ & $\delta_Q$ & $A$ & $B$ & $a$ & $\kappa$ & $\tau$ &$\varepsilon_a$ &$\varepsilon_p$ \\
\hline
$2.0$ & $0.5$ & $0.01$ & $0.01$ & $0.1$ & $0.1$ & $2.5$ & $2.0$ & $10.0$ & $0.2$ & $2000$ & $0.0$ &$0.355$\\
\hline
\end{tabular}
}
\label{tab:1}
\end{table}

\begin{backmatter}
\bmsection{Funding}This will be populated automatically during production.
\bmsection{Acknowledgments} S.R. was supported by UKRI Grant No. EP/Y027531/1. The work was partly supported by the French Renatech network of cleanroom facilities.

\bmsection{Disclosures} The authors declare no conflicts of interest.


\bmsection{Supplemental document} n/a
\end{backmatter}

\bibliography{POLAR}

\begin{thebibliography}{10}
\newcommand{\enquote}[1]{``#1''}

\bibitem{McMahonNRP23}
P.~L. McMahon, {\protect\JournalTitle{Nature Reviews Physics}} \textbf{5}, 717 (2023).

\bibitem{Zhao22}
L.~Zhao, \emph{Vector Dissipative Solitons} (Springer International Publishing, 2022), pp. 105--130.

\bibitem{ScirePRL03}
A.~Scir\`e, J.~Mulet, C.~R. Mirasso, \emph{et~al.}, {\protect\JournalTitle{Phys. Rev. Lett.}} \textbf{90}, 113901 (2003).

\bibitem{KrupaO17}
K.~Krupa, K.~Nithyanandan, and P.~Grelu, {\protect\JournalTitle{Optica}} \textbf{4}, 1239 (2017).

\bibitem{XuNC21}
G.~Xu, A.~U. Nielsen, B.~Garbin, \emph{et~al.}, {\protect\JournalTitle{Nat. Commun.}} \textbf{12} (2021).

\bibitem{XuOL22}
G.~Xu, L.~Hill, J.~Fatome, \emph{et~al.}, {\protect\JournalTitle{Optics Letters}} \textbf{47}, 1486 (2022).

\bibitem{HillCP24}
L.~Hill, E.-M. Hirmer, G.~Campbell, \emph{et~al.}, {\protect\JournalTitle{Communications Physics}} \textbf{7} (2024).

\bibitem{LuoOE20}
Y.~Luo, Y.~Xiang, P.~P. Shum, \emph{et~al.}, {\protect\JournalTitle{Opt. Express}} \textbf{28}, 4216 (2020).

\bibitem{HuangAPN23}
Z.~Huang, S.~Sergeyev, Q.~Wang, \emph{et~al.}, {\protect\JournalTitle{Advanced Photonics Nexus}} \textbf{2} (2023).

\bibitem{VladimirovOL19}
A.~G. Vladimirov, K.~Panajotov, and M.~Tlidi, {\protect\JournalTitle{Opt. Lett.}} \textbf{45}, 252 (2019).

\bibitem{MarconiNP15}
M.~Marconi, J.~Javaloyes, S.~Barland, \emph{et~al.}, {\protect\JournalTitle{Nature Photonics}} \textbf{9}, 450 (2015).

\bibitem{GarbinNC15}
B.~Garbin, J.~Javaloyes, G.~Tissoni, and S.~Barland, {\protect\JournalTitle{Nat. Commun.}} \textbf{6} (2015).

\bibitem{Yanchuk2019}
S.~Yanchuk, S.~Ruschel, J.~Sieber, and M.~Wolfrum, {\protect\JournalTitle{Phys. Rev. Lett.}} \textbf{123}, 053901 (2019).

\bibitem{Pammi2019}
V.~A. Pammi, K.~Alfaro-Bittner, M.~G. Clerc, and S.~Barbay, {\protect\JournalTitle{IEEE J. Sel. Topics Quantum Electron.}} \textbf{26}, 1 (2019).

\bibitem{prucnal2017neuromorphic}
P.~R. Prucnal and B.~J. Shastri, \emph{Neuromorphic photonics} (CRC press, 2017).

\bibitem{Shastri2021}
B.~J. Shastri, A.~N. Tait, T.~{Ferreira de Lima}, \emph{et~al.}, {\protect\JournalTitle{Nat. Photonics}} \textbf{15}, 102 (2021).

\bibitem{ZhangOL19}
Y.~Zhang, S.~Xiang, X.~Guo, \emph{et~al.}, {\protect\JournalTitle{Opt. Lett.}} \textbf{44}, 1548 (2019).

\bibitem{XiangOL20}
S.~Xiang, Z.~Ren, Y.~Zhang, \emph{et~al.}, {\protect\JournalTitle{Optics Letters}} \textbf{45}, 1104 (2020).

\bibitem{ScireOL02}
A.~Scirè, J.~Mulet, C.~R. Mirasso, and M.~San~Miguel, {\protect\JournalTitle{Opt. Lett.}} \textbf{27}, 391 (2002).

\bibitem{Barbay2011}
S.~Barbay, R.~Kuszelewicz, and A.~M. Yacomotti, {\protect\JournalTitle{Opt. Lett.}} \textbf{36}, 4476 (2011).

\bibitem{Selmi2014}
F.~Selmi, R.~Braive, G.~Beaudoin, \emph{et~al.}, {\protect\JournalTitle{Phys. Rev. Lett.}} \textbf{112}, 1 (2014).

\bibitem{Terrien2018a}
S.~Terrien, B.~Krauskopf, N.~G.~R. Broderick, \emph{et~al.}, {\protect\JournalTitle{Opt. Lett.}} \textbf{43}, 3013 (2018).

\bibitem{Ruschel2020limits}
S.~Ruschel, B.~Krauskopf, and N.~G. Broderick, {\protect\JournalTitle{Chaos}} \textbf{30}, 093101 (2020).

\bibitem{PammiThese21}
V.~A. Pammi, \enquote{{Photonic computing with coupled spiking micropillars and extreme event prediction in microcavity lasers},} Theses, {Universit{\'e} Paris-Saclay} (2021).

\bibitem{Terrien2019}
S.~Terrien, V.~A. Pammi, N.~G.~R. Broderick, \emph{et~al.}, {\protect\JournalTitle{Phys. Rev. Research}} \textbf{2}, 023012 (2020).

\bibitem{TerrienChaos23}
S.~Terrien, B.~Krauskopf, N.~G.~R. Broderick, \emph{et~al.}, {\protect\JournalTitle{Chaos}} \textbf{33} (2023).

\bibitem{TerrienPRE21}
S.~Terrien, V.~A. Pammi, B.~Krauskopf, \emph{et~al.}, {\protect\JournalTitle{Phys. Rev. E}} \textbf{103}, 012210 (2021).

\bibitem{MartinRegaladoIEEEJQE97}
J.~Martin-Regalado, F.~Prati, M.~San~Miguel, and N.~Abraham, {\protect\JournalTitle{IEEE Journal of Quantum Electronics}} \textbf{33}, 765 (1997).

\bibitem{ExterPRL98}
M.~P. van Exter, A.~Al-Remawi, and J.~P. Woerdman, {\protect\JournalTitle{Physical Review Letters}} \textbf{80}, 4875 (1998).

\bibitem{VanDerSande2006}
G.~{Van Der Sande}, M.~Peeters, I.~Veretennicoff, \emph{et~al.}, {\protect\JournalTitle{IEEE J. Quantum Electron.}} \textbf{42}, 898 (2006).

\bibitem{Panayotov2013}
K.~Panayotov and F.~Prati, \emph{Polarization Dynamics of VCSELs} (Springer, 2013), vol. 166 of \emph{Springer Series in Optical Sciences}, pp. 181--232.

\bibitem{Yamada1993}
M.~Yamada, {\protect\JournalTitle{IEEE J. Quantum Electron.}} \textbf{29}, 1330 (1993).

\bibitem{Dubbeldam1999}
J.~L. Dubbeldam and B.~Krauskopf, {\protect\JournalTitle{Opt. Commun.}} \textbf{159}, 325 (1999).

\bibitem{Krauskopf2011}
B.~Krauskopf and J.~Walker, \emph{Bifurcation study of a semiconductor laser with saturable absorber and delayed optical feedback} (Wiley-VCH Verlag, Germany, 2011).

\bibitem{Terrien2018}
S.~Terrien, B.~Krauskopf, N.~G. Broderick, \emph{et~al.}, {\protect\JournalTitle{Phys. Rev. A}} \textbf{98}, 1 (2018).

\bibitem{Terrien2017a}
S.~Terrien, B.~Krauskopf, N.~G. Broderick, \emph{et~al.}, {\protect\JournalTitle{Phys. Rev. A}} \textbf{96}, 1 (2017).

\bibitem{Sieber2014}
J.~Sieber, K.~Engelborghs, T.~Luzyanina, \emph{et~al.}, {\protect\JournalTitle{arXiv}} \textbf{1406.7144} (2014).

\end{thebibliography}

\bibliographyfullrefs{POLAR}


\ifthenelse{\equal{\journalref}{aop}}{%
\section*{Author Biographies}
\begingroup
\setlength\intextsep{0pt}
\begin{minipage}[t][6.3cm][t]{1.0\textwidth} 
  \begin{wrapfigure}{L}{0.25\textwidth}
    \includegraphics[width=0.25\textwidth]{john_smith.eps}
  \end{wrapfigure}
  \noindent
  {\bfseries John Smith} received his BSc (Mathematics) in 2000 from The University of Maryland. His research interests include lasers and optics.
\end{minipage}
\begin{minipage}{1.0\textwidth}
  \begin{wrapfigure}{L}{0.25\textwidth}
    \includegraphics[width=0.25\textwidth]{alice_smith.eps}
  \end{wrapfigure}
  \noindent
  {\bfseries Alice Smith} also received her BSc (Mathematics) in 2000 from The University of Maryland. Her research interests also include lasers and optics.
\end{minipage}
\endgroup
}{}

\end{document}